\documentclass[12pt]{article}
\usepackage{graphicx}
\begin{document} 

\centerline{\it Draft}

\bigskip
\centerline{\bf
Non-equilibrium and Irreversible Simulation}

\centerline {\bf of Competition among Languages}

\bigskip

\noindent
D. Stauffer$^1$, C. Schulze$^1$, F.W.S. Lima$^2$, 
S. Wichmann$^3$, and S. Solomon$^4$

\bigskip
\noindent
$^1$ Institute for Theoretical Physics, 

Cologne University, D-50923 K\"oln, Euroland

\noindent
$^2$ Departamento de F\'{\i}sica,

Universidade Federal do Piau\'{\i}, 57072-970 Teresina - PI, Brazil

\noindent
$^3$ Department of Linguistics, Max Planck Institute for Evolutionary 

Anthropology, Deutscher Platz 6, D-04103 Leipzig, Germany

\noindent
$^4$ Racah Institute of Physics, 

Hebrew University, IL-91904 Jerusalem, Israel

\bigskip
\centerline{e-mail: stauffer@thp.uni-koeln.de}
\bigskip

Abstract: The bit-string model of Schulze and Stauffer (2005)
is applied to non-equilibrium situations and then gives better
agreement with the empirical distribution of language sizes.
Here the size is the number of people having this language as
mother tongue. In contrast, when equilibrium is combined with
irreversible mutations of languages, one language always
dominates and is spoken by at least 80 percent of the population.


\section{Introduction}
Computer simulations of languages have a long tradition 
\cite{Cangelosi,Nettle}, particularly for the
learning of one language \cite{Nowak,Baronchelli}. More recent is
the simulation of the extinction of old and the emergence of
new languages 
\cite{Abrams,Patriarca,Mira,Pinasco,Wang,Kosmidis,Schwammle,Oliveira}.
To explain the coexistence of 
$10^4$ present human languages, the bit-string model of Schulze and Stauffer 
\cite{Schulze}, continued in \cite{Tesileanu}, is particularly 
convenient for simulation since with 8 or 16 bits it simulates $2^8 = 256$ or 
$2^{16} = 65536$ different languages (or different grammars). Also
the model of de Oliveira et al \cite{Oliveira} allows for numerous 
languages, but does not allow for the distinction of equilibrium
from non-equilibrium which the present note concentrates on.

One of the open questions of linguistics is, whether the present
distribution of human languages (e.g. \cite{Wichmann}) is 
similar to those of ancient past and distant future, or whether
it merely presents a transition between a past multitude of
languages and a future paucity: Every ten days on average a 
human language dies out, and in Brazil already half the 
indigeneous languages were replaced during the last half 
millenium by Portuguese. On the other hand, national languages 
like Hebrew for Israel were resurrected, and Francophones in 
Quebec fight for the survival of their French as official 
language in all of Quebec; a continuous, albeit slow, addition
of languages takes place by the drifting apart of dialects.

We leave it to others to separate languages from dialects, and
follow standard statistics \cite{Grimes}, as cited in \cite{Wichmann}, for
the present language sizes. The size $s$ of a language is defined
as the number of people speaking it as mother language, and 
varies from $10^9$ for Mandarin Chinese down to 1 for the last 
surviving speaker of a dying language. The number $n_s$ of 
languages with $s$ speakers each follows roughly a log-normal
distribution, 
$$\log[n_s/n_{s_{\max}}] \propto -[\log(s) - \log(s_{\max})]^2 \quad,$$
where $s_{\max}$ is the language size which appears most often.
In a double-logarithmic plot of log $n_s$ versus log $s$ such
data follow a parabola with a maximum at $s_{\max}$. Sutherland
\cite{Sutherland} pointed out that for small sizes below 10 the observed 
number of languages is higher than the log-normal distribution,
and our Fig.1 shows these data.

Past simulations of the bit-string model of Schulze and Stauffer
\cite{Schulze} showed this desired log-normal distribution with upward
deviations for the smallest sizes, but $s$ seldomly exceeded 100 
or 1000. The model of \cite{Oliveira} gave much larger 
$s$ but in our simulations gave two power laws instead of one
log-normal distribution. We now try two modifications, of which
one worked, in the next section.

\begin{figure}[hbt]
\begin{center}
\includegraphics[angle=-90,scale=0.5]{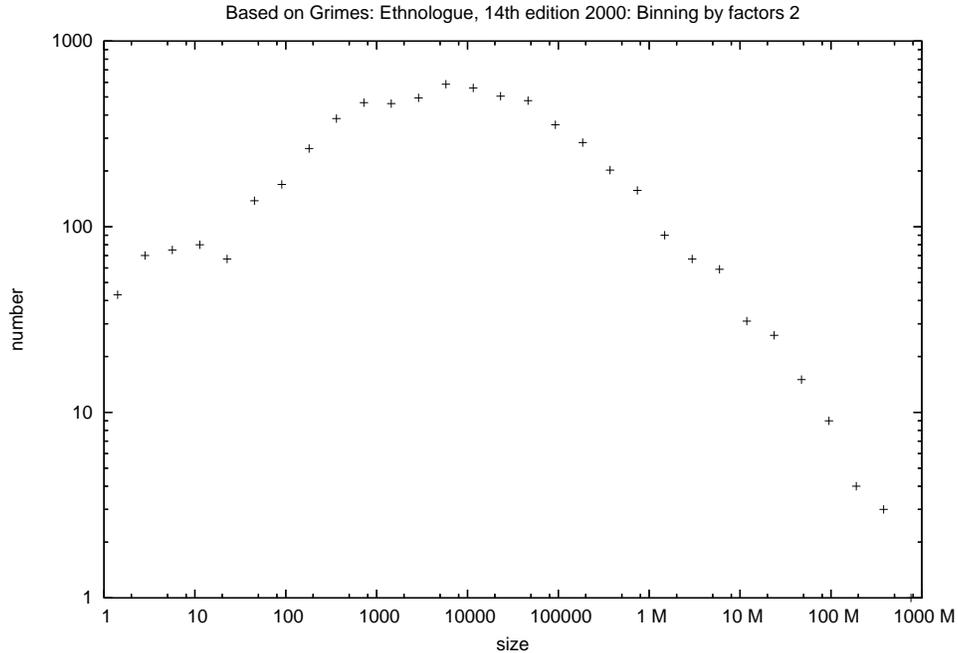}
\end{center}
\caption{Size histogram $n_s$ for human languages. We binned the sizes $s$ by 
factors of two; for example, all languages with sizes from 64 to 127 were
summed up into one data point. Binning by factors ten gives smoother data
\cite{Sutherland}. In these double-logarithmic plots a log-normal distribution
forms a parabola.
}
\end{figure}

\begin{figure}[hbt]
\begin{center}
\includegraphics[angle=-90,scale=0.5]{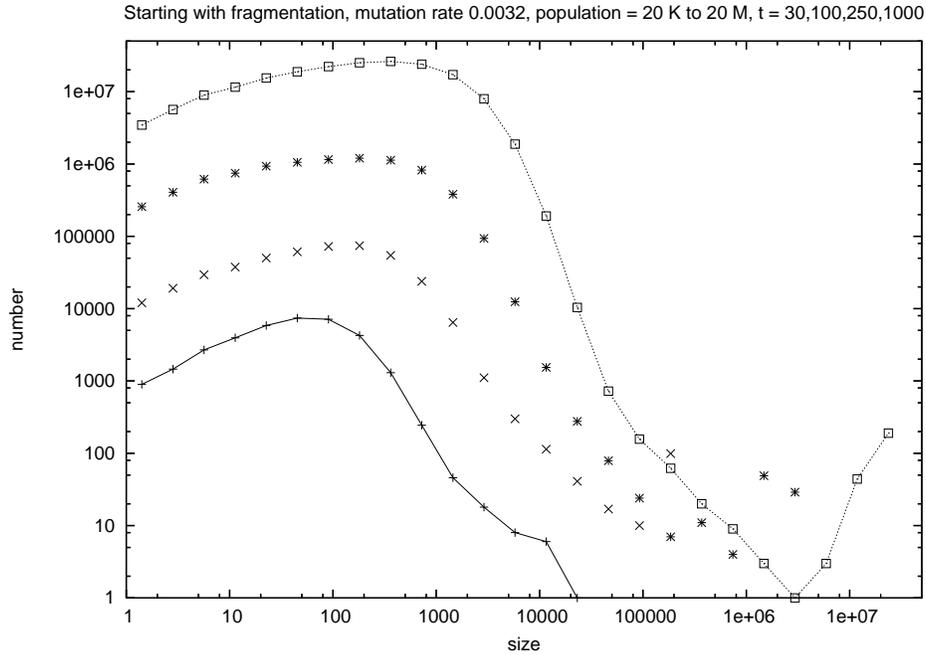}
\end{center}
\caption{Double-logarithmic plot of $n_s$ for population sizes from 20 thousand
to 20 million (from left to right), various times $t$ (= numbers of iterations),
and a mutation rate of 0.0002 per bit. We start with a random (fragmented) 
distribution of languages. Here and later we usually sum over ten simulations.
}
\end{figure}

\begin{figure}[hbt]
\begin{center}
\includegraphics[angle=-90,scale=0.5]{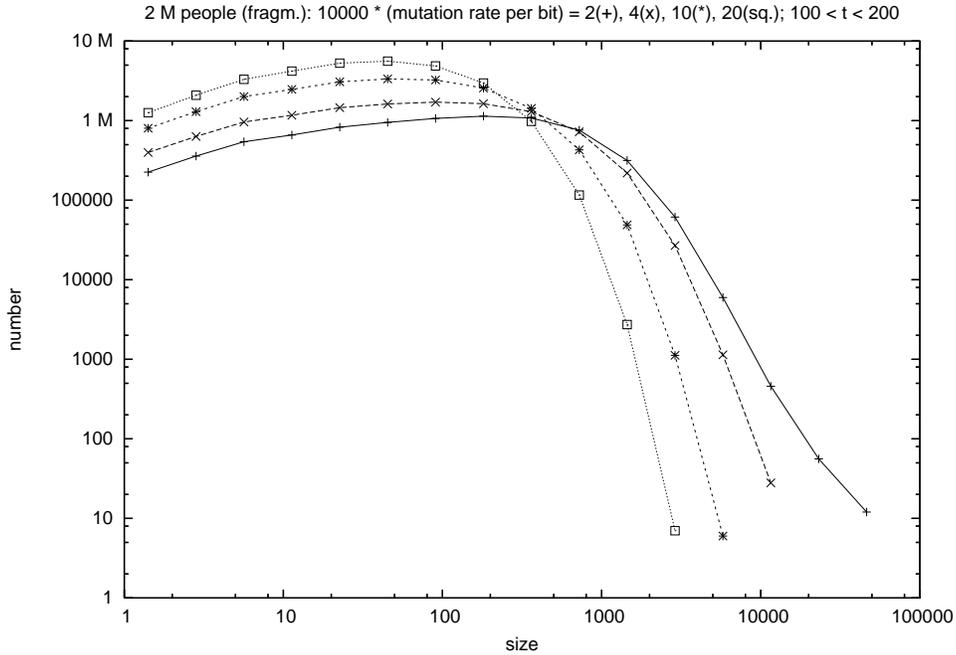}
\end{center}
\caption{Double-logarithmic plot of $n_s$ for 2 million people, summing up all
100 times between
101 and 200 iterations, and mutation rates per bit of 0.0002, 0.0004, 0.0010,
0.0020. For increasing mutation rate the distribution shifts to the left.
Again fragmented start.
}
\end{figure}

\begin{figure}[hbt]
\begin{center}
\includegraphics[angle=-90,scale=0.5]{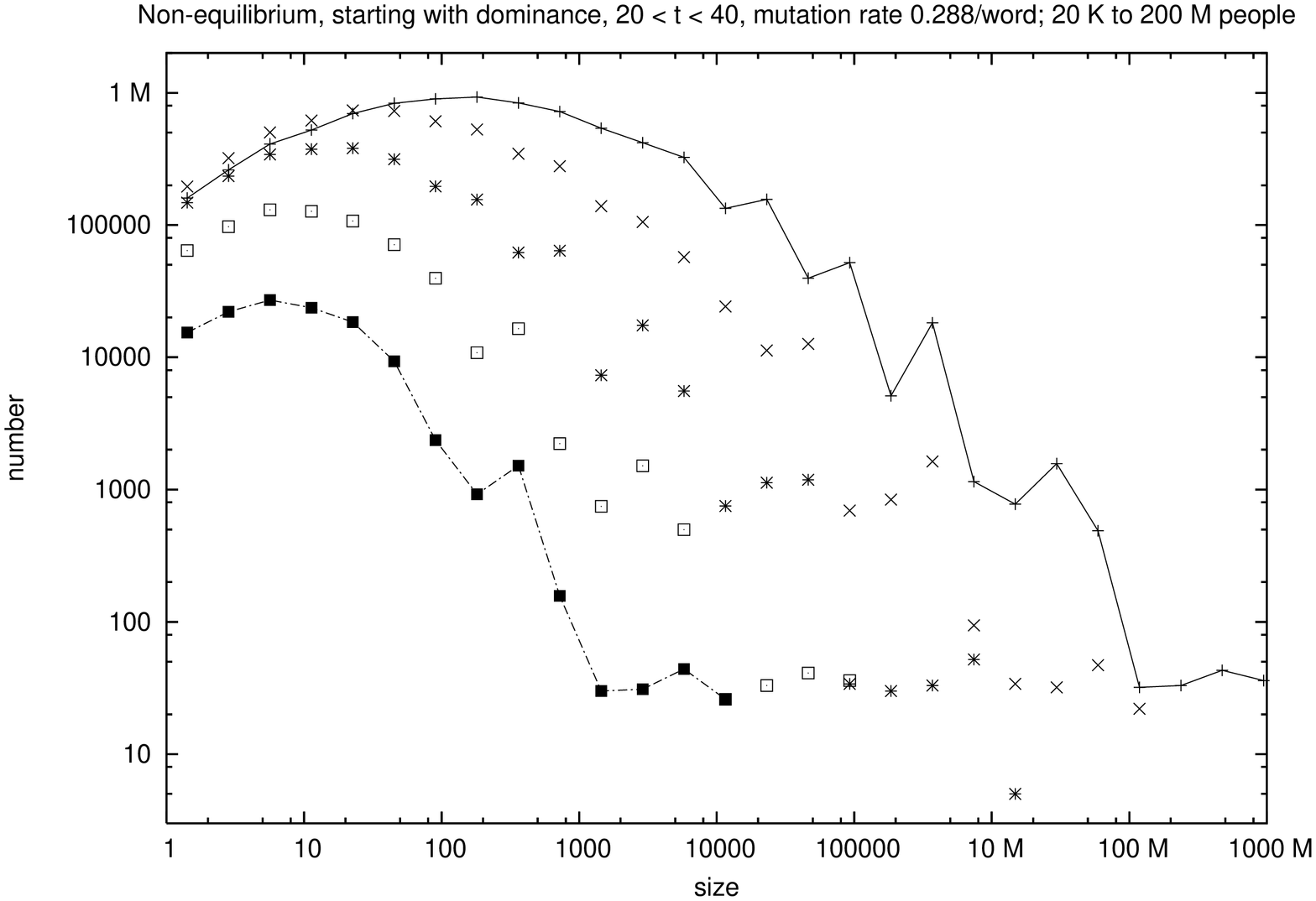}
\end{center}
\caption{Double-logarithmic plot of $n_s$ for 0.02, 0.2, 2, 20 and 200 million
people from left to right, summed over all 20 iterations between 21 and 40, 
mutation rate 0.018 per bit.
In contrast to Figs.2 and 3 we start here with everybody speaking the same 
language.
}
\end{figure}

\section{Computer Model}

The model symbolizes each language as a string of 16 bits, each
of which can be up (1) or down (0) and may represent some 
important aspect of the grammar. Simulations with less or more
bits, or with more than two choices for each grammatical feature,
were also made but gave qualitatively similar results. Languages
are defined as different if they differ in at least one bit.

At birth, the child adopts the mother language, apart from a
possible modification ("mutation", reversal of one bit). Adults
speaking a rare language may switch to a more widespread language.
After a few hundred time steps 
a dynamic equilibrium is established where the deaths 
roughly cancel the births and the size distribution no longer
changes systematically. In this equilibrium we either have a
dominance of one language spoken by at least 3/4 of the total
population, or the fragmentation of the population into up
to 65536 different languages of roughly equal sizes. The
choice depends on parameters and initial conditions. More details
are given in the appendix. 

One time step or iteration means that each individual 
individual is updated once: switching languages, giving birth, 
dying. In a comparison with reality, a "person" may correspond to a 
whole line of ancestors and offspring since one iteration may correspond 
to several centuries for the present mutation rates. A proper scaling 
of probabilities in order to allow one iteration to correspond to 
arbitrary time intervals still needs to be done.

Modification A is very simple: "Irreversibility". Bits may be 
changed from 0 to 1 but never from 1 to 0. In this case, language
1111111111111111 (i.e. all 16 bits set to 1) plays a special
role since it never changes. Languages rarely
change from state A to state B and then back again. For some phenomena this
may happen, e.g., word order, but irreversibility is more common, for
instance as concerns phonological changes.

Modification B is "Nonequilibrium with noise". We look
not at the late times of equilibrium but at the earlier times,
and average over an extended time interval which includes the
transition from fragmentation to dominance, or from dominance 
to fragmentation. Now we have one more important parameter, the
observation time, and we adjust it such that the results approximate a
log-normal distribution. 

Noise is included as random multiplicative: At each iteration, 
each language size $s$ is ten times multiplied by a factor
selected randomly between 0.9 and 1.1, with different factors for
each of the ten different multiplications. This noise 
incorporates changing pressures to use various languages; for
example if the present authors would have reported this work
a century ago they presumably would not have written it in
English. The noise also approximates the changes in birth 
rates, migrations, ethnic intermarriage, etc. due to external influences.

In both cases we select the initial population such that apart 
from minor fluctuations it agrees with the equilibrium 
population.

\section{Results}

Modification A (irreversibility) always resulted in dominance
even if we started with fragmentation. Depending on the mutation
probability, the end result was everybody speaking the same 
language (usually the one with 16 bits set to 1), or with between
80 and 100 percent of the people speaking one language and 
the others distributed among many other languages. None of these
two choices is what we want: Mandarin Chinese is spoken by only one sixth 
of the human population.

Modification B (non-equilibrium with noise) gave Fig.2 if we 
start with fragmentation: The whole population is randomly 
distributed among the 65536 languages. For increasing population
size we get larger languages and better results, but the shape of
the curves does not change much. Figure 3 shows for two million
people the rather minor effects of changing mutation rates,
between 0.0002 and 0.002 per bit or 0.0032 and 0.032 per 
bit-string. The tails at the right end depend strongly on when we stopped
the simulation and are anyhow far below the numbers (relative to the maximum) 
seen in reality, Fig.1.

If we start with dominance, i.e. everybody speaks
language zero,  we need a larger mutation
rate to get fragmentation; Fig.4 shows again log-normal 
distributions for various total population sizes, but now the 
left part near the maximum looks like the corresponding right part near the 
maximum. Reality, Fig.1, shows enhanced numbers in the left part, and this
left-right asymmetry is visible in Figs.2 and 3, not in Fig.4. Thus while
Figs.2 and 3 may correspond to the present shrinking of 
language diversity, Fig.4 may average over the times shortly 
before and after the biblical story of the Tower of Babel.

\section{Conclusion}

Figure 2 for modification B shows that with non-equilibrium
and random multiplicative noise we got the desired roughly
log-normal distribution with an enhanced number of languages for 
small sizes, just as in present reality, Fig.1. This is an
argument in favor of regarding the present language sizes as
a transient phenomenon between past fragmentation and future
dominance. But since it is only a computer model, it does not 
prove that in the future everybody will speak Mandarin Chinese and its 
mutants.

\bigskip
We thank J. Jost for initiating this cooperation, and M. Cysouw for suggestions 
which lead to modification A.

\section{Appendix: The old bit-string model}

Our simulation model in general is based on three probabilities $p,q,r$ for
change, for transfer, and for flight from small languages. A language (perhaps 
better: a grammar) is defined as a chain of $F$ features $(4 \le F \le 64)$ each
of which can take one of $Q$ different values $1,2,\dots,Q$ with $2 \le Q \le
10$. The binary case $Q=2$ allows memory-saving representation as bit-strings,
particularly if $F = 8, 16, 32$ or 64. In the present paper we use only $F=16,\;
Q=2$, storing the whole "language" in one two-byte computer word. When a child 
is born, with probability $p$ its language differs from that of the mother
(fathers are assumed not to help in rearing children and are thus neglected) on
one randomly selected position where the bit is changed with probability $q$ 
and is taken from the corresponding bit of a randomly selected other person
with probability $1-q$. In the present paper we set the transfer probability 
$q$ to zero: No language learns from other languages. 

Finally, speakers of 
small languages switch with a probability $r$ to the language of a randomly 
selected other person (which usually is a widespread language); this $r$ is
quadratic in the fraction of people speaking a language since a language is 
mostly used for communication between two people. If $x_i$ is the fraction of 
people  speaking language $i$, then $r = (1-x_i)^2, \; r = 1 - x_i^2$ and
$r = 0.9 (1-x_i^2)$ have been used; the present paper uses $1-x_i^2$ if we 
start with everybody speaking the same language, and $(1-x_i)^2$ if the initial
population is distributed randomly among the 65536 possible languages. 

Each person gives birth to one child per iteration, and dies with a Verhulst
probability proportional to the current total number $N(t)$ of people, due to 
lack of space and food. In the present paper we start with the same population 
which for the given Verhulst probability is already the equilibrium population
$N_0$. If instead one starts with only one person, the flight probability $r$ 
is reduced by a factor $N(t)/N_0$ since for low population densities the 
selection pressure on languages is weaker \cite{Nettle}. We averaged the 
numbers $n_s$ of language sizes over the second half of the simulation.

A complete program with description is published in \cite{book}; the
present programs are available by e-mail as language35.f and language36.f.

\end{document}